\documentclass{larnacaconf}

\def\pd#1#2{\frac{\partial#1}{\partial#2}}
\def\IK{\relax{\rm l\kern-.18 em K}}                                    
\begin{document}

\FirstPageHeading{Carinena}

\ShortArticleName{Two important examples of nonlinear  oscillator}

\ArticleName{Two important examples of nonlinear  oscillators}

\Author{Jos\'e F.  CARI\~NENA~$^\dag$, Manuel F. RA\~NADA~$^\dag$ and Mariano SANTANDER~$^\ddag$}
\AuthorNameForHeading{J.F. Cari\~nena,  M.F. Ra\~nada  and M. Santander}
\AuthorNameForContents{CARI\~NENA J.F., RA\~NADA M.F. and SANTANDER M.}
\ArticleNameForContents{Two important examples of nonlinear  oscillators}

\Address{$^\dag$~Departamento de F\'{\i}sica Te\'orica, Facultad de Ciencias, 
\\
Universidad de Zaragoza,
50009 Zaragoza, Spain} 
\EmailD{jfc@unizar.es,\ mfran@unizar.es}

\Address{$^\ddag$~ Departamento de F\'{\i}sica Te\'orica, Facultad de Ciencias \\
  Universidad de Valladolid,  47011 Valladolid, Spain}
\EmailD{santander@fta.uva.es}


\Abstract{We discuss a classical   nonlinear  oscillator,
which is  
 proved to be a superintegrable system for which 
 the bounded motions are quasiperiodic oscillations and the
unbounded (scattering) motions are represented by hyperbolic functions.
This oscillator  can be seen as a 
position-dependent mass system  and we show a natural quantization 
prescription 
admitting a factorization with  shape invariance for the $n=1$ case,
and then the energy spectrum is found.
Other isochronous systems which can also be considered as a generalization
 of the harmonic oscillator and admit a nonstandard Lagrangian description
are also discussed.
}

\section{Introduction}

The harmonic oscillator is a system playing a privileged r\^ole both in 
classical and quantum mechanics. It is almost ubiquitous in 
Physics and appears 
in many physical applications running from condensed matter to semiconductors 
(see e.g. \cite{YD}
for references to such problems).
The dynamical evolution of the classical system in one dimension is given by 
$$\frac {dq}{dt}=v\,,\qquad 
\frac {dv}{dt}=-\omega^2\, q\ ,
$$
and admits a Lagrangian formulation with 
$L= (1/2) \left(v^2-\omega^2\, q^2\right)$, 
the  general solution of the equations of motion being
$$q=q_0\, \cos \omega t-\frac{v_0}\omega \sin\omega t=A\cos(\omega t+\varphi)$$
and therefore the solutions are periodic with angular frequency $\omega$,
while $A$ and $\varphi$ are arbitrary. This is the main characteristic of
 the classical system. As far as the quantum system is concerned,
 the eigenvalues of the Hamiltonian, which is  given 
by 
$H=(1/2)(p^2+\omega^2\, q^2)$, are equally spaced. We should  also remark
  that the natural extensions to two dimensions, given by 
$$
   H = {1\over 2}\,(p_x^2 + p_y^2)
     + {1\over 2}\,(\omega_1^2\, x^2 + \omega_2^2\, y^2)
  $$
admits two constants of motion in involution,
$I_1=E_x=\frac 12\,(p_x^2+\omega_1^2\, x^2)$, $I_2=E_y=\frac 12\,(p_v^2+\omega_2^2\, y^2) $, 
 and therefore it is completely integrable in the sense of Liouville.
Moreover it has been proved that, when $\omega_1$ and $\omega_2$ 
are rationally related, i.e.
$\omega_1 = n_1\,{\omega_0}$, $
\omega_2 = n_2\,{\omega_0}$, with $n_1,n_2\in \mathbb{N}$, there exist a new constant of motion and  the system 
is superintegrable. Actually the complex function, $
   J = K_x^{n_2}\,(K_y^{*})^{n_1} $ with 
$K_x = p_x + i\, n_1\,{\omega_0}\, x$ and 
$K_y = p_y + i\, n_2\,{\omega_0}\,y$, is a constant of the motion.

Our aim is to comment on some possible generalizations of this system from the perspective of the theory of the symmetry, i.e. trying to preserve the fundamental symmetry properties.

\section{A position-dependent mass nonlinear  oscillator}

A often used generalization was proposed by   Mathews and Lakshmanan  \cite{Carinena:MaL74,Carinena:LaRa03}  as a one-dimensional analogue of 
some models of quantum field theory \cite{Carinena:DeSS69,Carinena:NiW72}. 
It is described by a Lagrangian
\begin{equation}
  L  = \frac{1}{2}\,\Bigl(\frac{1}{1 + \lambda\,x^2} \Bigr)\,(\dot{x}^2 - 
\alpha^2\,x^2)\ ,
\end{equation}
 which can be  considered as an oscillator
with a position-dependent effective mass $m=(1 + \lambda\,x^2)^{-1} $
(see e.g. \cite{Carinena:Le95,Carinena:KK03} and references therein). It was proved 
that the general solution is also $q(t)=A\cos(\omega t+\varphi)$, but now
the amplitude $A$ depends on  the frequency. More explicitly
$\omega^2\,(1 + \lambda\,A^2)  = \alpha^2$. Note also that this Lagrangian is of mechanical type, the kinetic term being invariant under the tangent lift of the vector field 
$$
  X_x(\lambda) = \sqrt{\,1+\lambda\,x^2\,}\,\,\pd{}{x}  \,.
$$

It was recently shown in \cite{Carinena:CRSS04} that there is a generalization to $n$ dimensions preserving the symmetry characteristics. In particular the 
 two-dimensional generalization studied in  \cite{Carinena:CRSS04} was given by the Lagrangian 
\begin{equation}
  L(\lambda) = \frac{1}{2}\,\Bigl(\frac{1}{1 + \lambda\,r^2}\Bigr)\,
  \Bigl[\,v_x^2 + v_y^2 + \lambda\,(x v_y - y v_x)^2-\alpha^2\, r^2 \,\Bigr] 
\,,\quad
  r^2 = x^2+y^2,
\label{Carinena:Ln2}
\end{equation}
and it was shown to be not only integrable but also superintegrable. This is the only generalization to $n$ dimensions for which the kinetic term is 
a quadratic function in the velocities that is 
invariant under rotations and 
 under the two vector fields generalising the symmetries of the one-dimensional model, i.e.
$$  X_1(\lambda) = \sqrt{\,1+\lambda\,r^2\,}\,\,\pd{}{x}  \,,\qquad
   X_2(\lambda) = \sqrt{\,1+\lambda\,r^2\,}\,\,\pd{}{y}  \,.
$$

This is valid for any value of $\lambda$. However, when $\lambda<0$,
$\lambda=-\,|\lambda|$, this function  has 
a singularity
at $1 -\,|\lambda|\,r^2=0$ and we restrict our 
 dynamics
to the interior of the circle
$x^2+y^2<1/|\lambda|$.

These two vector fields close with the generator of rotations, 
$$ X_J   = x\,\pd{}{y} - y\,\pd{}{x}\,, $$ 
on a Lie algebra 
$$ [X_1(\lambda)\,,X_2(\lambda)] = {\lambda}\,X_J \,,{\quad}
  [X_1(\lambda)\,,X_J] =      X_2(\lambda)   \,,{\quad}
  [X_2(\lambda)\,,X_J]=    -\,X_1(\lambda)   \,.
$$
which  is isomorphic either to $SO(3,\mathbb{R})$, when $\lambda>0$, or 
to $SO(2,1)$, 
when $\lambda<0$, or finally  to the Euclidean group in two dimensions when $\lambda=0$.

The important property shown in \cite{Carinena:CRSS04} is that this 
bidimensional nonlinear harmonic oscillator is completely integrable,
because one can show that, 
if $K_1$ and  $K_2$ are the  functions
$$  K_1  = P_1(\lambda) + {i\,}{\alpha}\ \frac{x}{\sqrt{\,1 + 
\lambda\,r^2\,}} \,,\quad
   K_2  = P_2(\lambda) + {i\,}{\alpha}\ \frac{y}{\sqrt{\,1 + 
\lambda\,r^2\,}} \,,
$$
with $$
  P_1(\lambda) 
           = \frac{v_x - \lambda\,J y}{\sqrt{\,1 + \lambda\,r^2\,}}\,,\quad 
  P_2(\lambda) 
           = \frac{v_y + \lambda\,J x}{\sqrt{\,1 + \lambda\,r^2\,}}\,, \quad J=x v_y - y v_x\,,
$$
then the complex functions  $K_{ij}$ defined as
$  K_{ij} = K_i\,K_j^*$,  $ i,j=1,2$,
are constants of motion. In fact the 
 time-evolution of the functions $K_1$ and $K_2$ is
$$   \frac{d}{d t}\,K_1 
    =  \frac{{i}{\alpha}}{1 + \lambda\,r^2}\, K_1     \,,\quad
  \frac{d}{d t}\,K_2  =
 \frac{{i}{\alpha}}{1 + \lambda\,r^2}\,K_2 \,,
$$ 
from which we see that the complex functions $K_{ij}$ are constants of the motion.  Therefore the system is superintegrable with the following 
first integrals of motion
$$
  I_1(\lambda) = |\,K_1\,|^2 \,,{\quad}
  I_2(\lambda) = |\,K_2\,|^2 \,,{\quad}  
  I_3 = \Im(K_{12}) = {\alpha}\,(x v_y - y v_x)  \,.
$$

The Legendre transformation for a two-dimensional Lagrangian system 
of mechanical type with kinetic term as in (\ref{Carinena:Ln2})
is given by
$$
  p_x = \frac{(1 + \lambda\,y^2) v_x - \lambda\,x y v_y}{1 + \lambda\,r^2} \,,\quad
  p_y = \frac{(1 + \lambda\,x^2) v_y - \lambda\,x y v_x}{1 + \lambda\,r^2} \,,
$$
(note that  $x p_y - y p_x=x v_y - y v_x$) and
the general expression for the corresponding  $\lambda$-dependent Hamiltonian is
\begin{equation}
  H(\lambda) = \frac{1}{2}\,\Bigl[\,p_x^2 + p_y^2 + \lambda\,(x 
p_x + y p_y)^2 \Bigr]
  + \frac{1}{2}\,{\alpha^2}\,V(x,y)\,,
\end{equation}
and hence the associated Hamilton--Jacobi equation is
\begin{equation}
   \left(\pd{S}{x}\right)^2 + \left(\pd{S}{y}\right)^2
  + \lambda\,\left(x\,\pd{S}{x} + y\,\pd{S}{y}\right)^2 + \alpha^2 \,
V(x,y) = 2\, E \,.
\end{equation}

This equation is not separable in $(x,y)$ coordinates, but it was shown in \cite{Carinena:CRSS04} that
 there exist three particular  systems of 
orthogonal coordinates,  and three particular families of associated
potentials, in which such  Hamiltonians  admit a Hamilton--Jacobi
separability. The first system of 
coordinates is given by 
\begin{equation}
(z_x,y) \,,\quad z_x = \frac{x}{\sqrt{\,1 + \lambda\,y^2\,}} \,,
\end{equation}
for which the Hamilton--Jacobi equation becomes:
$$  (1 + \lambda\,z_x^2)\left(\pd{S}{z_x}\right)^2
  + (1 + \lambda\,y^2)^2\left(\pd{S}{y}\right)^2 +\alpha^2\,(1 + \lambda\,y^2)V
  = 2 (1 + \lambda\,y^2) E\ ,
$$
and therefore the Hamilton--Jacobi equation is separable
if the potential $V(x,y)$ can be written on the form
\begin{equation}
   V = \frac{W_1(z_x)}{\,1 + \lambda\,y^2} + W_2(y)\ .
\end{equation}
The potential is therefore integrable with the following two quadratic
integrals of motion
\begin{eqnarray}
  I_1(\lambda) &=& (1 + \lambda\,r^2)p_x^2 + {\alpha^2} W_1(z_x) \,,\cr
  I_2(\lambda) &=& (1 + \lambda\,r^2)p_y^2 - \lambda\,J^2 +
  {\alpha^2} \left(W_2(y) - \frac{\lambda\,y^2}{\,1 + \lambda\,y^2}\,W_1(z_x)
\right).
{\nonumber}\end{eqnarray} 

In a similar way, one can see, using coordinates 
$(x,z_y)$ with $ z_y = {y}(1 + \lambda\,x^2)^{-1/2}$, that 
the 
Hamilton--Jacobi equation
is separable when the potential $V(x,y)$ is 
 of the form
\begin{equation}
   V = W_1(x) + \frac{W_2(z_y)}{\,1 + \lambda\,x^2}\ .
\end{equation}
and the potential is integrable with the following two quadratic
first integrals:
\begin{eqnarray}
  I_1(\lambda) &=& (1 + \lambda\,r^2)p_x^2 - \lambda\,J^2 +
  {\alpha^2} \Bigr(W_1(x) - \frac{\lambda\,x^2}{\,1 + \lambda\,x^2}\,W_1(z_y)\Bigl) \,,\cr
  I_2(\lambda) &=& (1 + \lambda\,r^2)p_y^2 + {\alpha^2} W_2(z_y) .
{\nonumber}\end{eqnarray}

Finally using  polar coordinates $(r,\phi)$ the 
Hamiltonian $H(\lambda)$ is written
\begin{equation}
  H(\lambda) = \frac{1}{2}\,\left[\,(1 + \lambda\,r^2)p_r^2 + 
\frac{p_\phi^2}{r^2} \right]
  + \frac{\alpha^2}{2}\,V(r,\phi)
\end{equation}
and the Hamilton--Jacobi equation is given by
$$
    (1 + \lambda\,r^2)\left(\pd{S}{r}\right)^2
  + \frac{1}{r^2} \left(\pd{S}{\phi}\right)^2 + \alpha^2\,V(r,\phi) = 2\, E  \,.
$$
 
Then the equation admits separability when the  potential $V$ is of the form
\begin{equation}
V = F(r) + {G(\phi)}/{r^2}\ .
\end{equation}

Such a  potential $V$ is integrable with the following two quadratic first
integrals:
\begin{eqnarray}
  I_1(\lambda)  &=&  (1 + \lambda\,r^2)p_r^2 + \frac{1-r^2}{r^2}\,p_\phi^2
  + \alpha^2 \left[\, F(r) + \frac{1-r^2}{r^2}\,G(\phi) 
\right] \,,\cr&&\cr
  I_2(\lambda)  &=&  p_\phi^2 + \alpha^2 G(\phi)      \,.
{\nonumber}\end{eqnarray}
Consequently, the potential
$$
  V  = \frac{\alpha^2}{2}\, \Bigl(\frac{x^2+y^2}{1 + \lambda\,(x^2+y^2)} \Bigr)
$$
is super-separable since it is separable in three different
systems of coordinates $(z_x,y)$, $(x,z_y)$, and $(r,\phi)$ because 
\begin{equation}
  V \! =\! \frac{\alpha^2}{2}\,\Bigl(\frac{1}{1+\lambda\,y^2}\Bigr)
  \Bigl[\frac{z_x^2}{1+\lambda\,z_x^2}  +  y^2 \Bigr]      
   \! =\! \frac{\alpha^2}{2}\,\Bigl(\frac{1}{1+\lambda\,x^2}\Bigr)
  \Bigl[ x^2  +  \frac{z_y^2}{1+\lambda\,z_y^2} \Bigr]   
  \!  =\!  \frac{\alpha^2}{2}\,\Bigl(\frac{r^2}{1+\lambda\,r^2}\Bigr)\,.{\nonumber}
\end{equation}

\section{The one-dimensional quantum nonlinear oscillator}

We now consider the quantum case and restrict ourselves to the one-dimensional case. The first problem is to define the
quantum operator describing the Hamiltonian of this position-dependent mass system, because the mass function and the momentum $P$ do not commute and this fact gives
rise to an ambiguity in the ordering of factors. It has recently been proposed to avoid the problem by modifying the Hilbert space of functions describing the system \cite{Carinena:CRS04}. More explicitly we can consider the measure 
$d\mu=(1 + \lambda\,x^2)^{-1/2}\, dx$, which is invariant under the vector field 
$  X_x(\lambda) = \sqrt{\,1+\lambda\,x^2\,}\,\partial/\partial{x}$,
for  then  the operator $P=-i\,\sqrt{\,1+\lambda\,x^2\,}\,\partial/\partial{x}$ is 
selfajdoint in the space  $L^2({\Bbb R},d\mu)$.  In the case of the nonlinear oscillator in which  we are interested we can consider the Hamiltonian operator
\begin{equation}
\widehat{H}_1 = -\frac 12\,(1 + \lambda\,x^2)\,\frac{d^2}{dx^2}
     -\frac 12\,\lambda\,x\,\frac{d}{dx} +
\frac{1}{2}\, \frac{\alpha^2\,x^2}{1 + \lambda\,x^2}  \,. 
\end{equation}
The spectral problem of such operator can be solved by means of algebraic 
techniques. We first remark that if   $\beta$ is such that
$\alpha^2=\beta(\beta+\lambda)$, then $\widehat{H}'_1=\widehat{H}_1-\beta/2$
 can be factorized as a product
$\widehat{H}'_1=A^\dag(\beta)\,A(\beta)$ and
\begin{equation}
  A  =  \dfrac 1{\sqrt 2}\left(\sqrt{1 + \lambda\,x^2}\ \dfrac{d}{dx} +\dfrac{\beta\,x}{\sqrt{1 + \lambda\,x^2}} 
\right)  \,,
\end{equation}
for which its adjoint operator is
\begin{equation}
A^\dag =\dfrac 1{\sqrt 2}\left( -\sqrt{1 + \lambda\,x^2}\ \dfrac{d}{dx}  +\dfrac{\beta\,x}{\sqrt{1 + \lambda\,x^2}} \right)   \,.
\end{equation}
The important point is that the partner Hamiltonian $\widehat{H}'_2=A(\beta)\,
A^\dag(\beta)$ is related to $\widehat{H}'_1$ by 
$ \widehat H'_2(\beta)  = \widehat H'_1(\beta_1) + R(\beta_1)$ with $\beta_1=f(\beta)$ and 
where $f$ and $R$ are the functions $f(\beta) = \beta-\lambda$
and
$R(\beta) = \beta + (1/2)$. Hamiltonians admitting such factorization 
\cite{Carinena:IH} and
related  to its s partner in  such a way are said to be  shape 
invariant and their 
 spectra and the corresponding eigenvectors can be found by using the method proposed by Gendenshte\"{\i}n \cite{Carinena:Ge83,Carinena:GK85} (see also \cite{Carinena:CR00} for a modern presentation based on the Riccati equation).
Therefore, as the quantum nonlinear oscillator is shape 
invariant, we can develop the method proposed in \cite{Carinena:Ge83,Carinena:GK85} for finding both
 the spectrum and the corresponding eigenvectors. The spectrum is given by 
\cite{Carinena:CRS04}:
$$E_n=n\,\beta-\frac {n^2}2\,\lambda+\frac 12 \beta\,.$$
The existence of a finite or infinite number of bound states depends up on the sign of $\lambda$  as also discussed in \cite{Carinena:CRS04}.

\section{Periodic motions and another nonlinear oscillator}
Another possible generalization of the harmonic oscillator 
would be to look for alternative isochronous systems. For instance
 one can consider a 
potential $$U(x)=\left\{\begin{array}{cc} U_1(x)&{\rm if\ } x<0\cr
U_2(x)&{\rm if\ } x>0\,,\end{array}\right.
$$
where  $U_2(x)$ is an 
increasing function and $U_1(x)$
is a decreasing function, and try to determine the explicit
 functions $U_1$ and $U_2$ in order to have an isochronous system. The 
problem of the determination of the potential when the period is known 
as a function of the energy was solved by Abel \cite{Carinena:Ab26}. When the potential 
is symmetric the solution is unique. Therefore the only symmetric potential
 giving rise to isochronous motions around the origin is the harmonic 
oscillator. The isotonic oscillator is also symmetric and isochronous,
but the origin is a singular point and not a minimum of the potential.
Other nonsymmetric potentials can be used, for instance 
a potential given by 
$$U_1(x)=\omega_1^2\, x^2\ ,\quad U_2(x)=\omega_2^2\, x^2\ .
$$

If we want to find more general solutions for the symmetric case we may 
consider Lagrangians of a nonstandard mechanical type, in which there is
 no potential term. These more general Lagrangians can also be relevant
 in other problems. For instance
another interesting oscillator-like system has recently been studied by 
Chandrasekar {\sl et al} \cite{Carinena:CSL04}. As mentioned in that
 paper the oscillator-like system  admits a Lagrangian formulation. We recall that there are systems admitting a Lagrangian formulation of a nonmechanical type. As an example
 we can consider 
 $Q={\Bbb R }$ as the configuration space   and the Lagrangian function \cite{Carinena:CRS04b}
 \begin{equation}
L(x,v)=(\alpha(x)\, v+U(x))^{-1}\,,
\end{equation}
which is singular in the zero level set of the function 
$\varphi(x,v)=\alpha(x)\, v+U(x)$. 

Then  the Euler--Lagrange equation is
$$
\alpha'(x)\,v+U'(x)-\alpha'(x)\,v=-
\frac{2\alpha(x)
[\alpha'(x) v^2+ U'(x) v+\alpha(x) a]}{\alpha(x) v+U(x)}\ ,
$$
where $v$ and $a$ denote the velocity and  the acceleration, respectively.

This is a conservative system the  equation of motion of which can be rewritten as 
$$
[\alpha(x)]^2\,\ddot x+\alpha(x)\,\alpha'(x)\,\dot x^2+\frac 32 \,\alpha(x)\, U'(x)\,\dot x+\frac 12 \,U(x)\,U'(x)=0\ .
$$

The energy  is given by 
$
E_L(x,v)=-
[2\,\alpha(x)\, v+U(x)][\alpha(x)\, v+U(x)]^{-2}$.
In
 particular, when $\alpha(x)=1$, the Lagrangian is 
$L(x,v)=[v+U(x)]^{-1}$ and
 the Euler--Lagrange equation reduces to
\begin{equation}
\ddot x+\frac 32 \, U'(x)\,\dot x+\frac 12 \,U(x)\,U'(x)=0
\end{equation}
and the energy function turns out to be 
$E_L(x,v)=-[2\, v+U(x)][ v+U(x)]^{-2}$.
When  $U(x)= k\, x^2$,  the equation is
$$
\ddot x+3\, k\,x\, \dot x+k^2\, x^3=0\ ,
$$
and the energy is $E_L=-[2\, v+k\,x^2][v+k\,x^2]^{-1}$.
It can be seen from the energy conservation law that the general solution is
$$x=\frac {2\, t}{k\, t^2-E}\ .
$$

The two-dimensional system described by 
$L(x,y,v_x,v_y)=[v_x+k_1\,x^2]^{-1}+[v_y+k_2\,y^2]^{-1}$
is superintegrable. Actually not only the energies of each degree of freedom are conserved but also the functions \cite{Carinena:CRS04b}
$$I_3=\frac x{v_x+k_1\,x^2}-\frac y{v_y+k_2\,y^2}\,,\quad I_4=\frac {k_2}{v_x+k_1\,x^2}+\frac {k_1}{v_y+k_2\,y^2}-\frac{k_1\,k_2\,x\,y}{(v_x+k_1\,x^2)(v_y+k_2\,y^2)}\ .
$$

Another example is that of a nonlinear oscillator for which we were looking. 
The following Lagrangian depending on the parameter $\omega$
\begin{equation}
L(x,v;\omega) =\frac 1{k\,v_x+k^2\,x^2+\omega^2}\ ,
\end{equation}
produces the nonlinear  Euler--Lagrange equation 
$$\ddot x+3\, k\, x\,\dot x+k^2\, x^2+\omega^2x=0\ ,$$
which is the nonlinear oscillator system recently studied
 by Chandrasekar {\sl et al} \cite{Carinena:CSL04}, 
and the energy is $E_L=-[2\,k\, v_x+k^2\,x^2+\omega^2][(k\, v_x+k^2\,x^2+\omega^2)]^{-2}$.
The general solution for the dynamics, which can be found from the energy conservation, is
$$x=\frac{\omega\,{\sqrt E}\,\sin (\omega t+\phi)}{1-k\, {\sqrt E}\,
\cos (\omega t+\phi)}\ .
$$

We have recently been able to prove \cite{Carinena:CRS04b} that in the 
rational case of the two-dimensional problem, for which 
$\omega_1=n_1\, \omega_0$ and $\omega_2=n_2\, \omega_0$,
the system is superintegrable as it was the case for the harmonic oscillator.
 To introduce the additional constants of motion we define 
\begin{equation}
\IK_1=\frac{v_x+k_1\,x^2+i\,n_1\, \omega_0\, x}{k_1\,v_x+k_1^2\,x^2+n_1^2\,\omega_0^2}\,,\qquad \IK_2=\frac{v_y+k_2\,y^2+i\,n_2\, \omega_0\, y}{k_2\,v_y+k_2^2\,y^2+
n_2^2\,\omega_0^2}\,,
\end{equation}
and then the complex function $\IK_1^{n_2}\,(\IK_2^*)^{n_1}$ is a constant of the motion.

In summary, not only position-dependent mass generalizations of the harmonic
oscillator can be interesting but there exist also systems described by
Lagrangians of non-mechanical type which preserve the property of
superintegrability for the harmonic oscillator with rationally related
frequencies. This example points out the importance of the study of 
such non-standard Lagrangians.

\LastPageEnding

\end{document}